# THE FNAL BOOSTER 2ND HARMONIC RF CAVITY*

R. Madrak[†], J. Dey, K. Duel, M. Kufer, J. Kuharik, A. Makarov, R. Padilla, W. Pellico, J. Reid, G. Romanov, M. Slabaugh, D. Sun, C. Y. Tan, I. Terechkine,
Fermilab, Batavia, IL, 60510, USA

*Abstract*

A second harmonic RF cavity which uses perpendicularly biased garnet for frequency tuning is currently being constructed for use in the Fermilab Booster. The cavity will operate at twice the fundamental RF frequency, from ~76 - 106 MHz, and will be turned on only during injection, and transition or extraction. Its main purpose is to reduce beam loss as required by Fermilab's Proton Improvement Plan (PIP). After three years of optimization and study, the cavity design has been finalized and all constituent parts have been received. We discuss the design aspects of the cavity and its associated systems, component testing, and status of the cavity construction.

## INTRODUCTION

The defining feature of this cavity is its on-axis tuner using aluminum doped garnet: National Magnetics AL-800. The tuning is achieved by sweeping the bias magnetic field, which is perpendicular to the RF magnetic field. This is different from many wideband cavities which use materials such as NiZn ferrite, where the ferrite is biased parallel to the RF magnetic field. Using garnet is desirable because the saturation magnetization is typically lower than in ferrite, and with a realistic magnetic system it can be biased to saturation where losses are lower. However, to maintain tunability, the bias must then be perpendicular to the RF magnetic field instead of parallel. If the tunability is sufficient, substantially higher shunt impedances can be attained.

Several cavities with a perpendicular bias [1] have been constructed and used operationally [2], but these have limited tuning range. Both TRIUMF/LANL [3] and the SSC Low Energy Booster [4] developed prototype cavities with large tuning ranges, but none of these cavities were ever used with beam.

## PURPOSE

It is well known that by flattening the bucket at injection, it is possible to increase the capture efficiency because of increased bucket area and a reduction in space charge density [5]. Although beam capture in Booster is already quite efficient, greater than 90% for $5.3 \times 10^{12}$ protons, there is still an activation problem due to beam loss. Therefore, even a gain in efficiency of a few percent can help mitigate this problem. This is the main motivation for the installation of a 2nd harmonic cavity in the Booster. The cavity will be turned on for approximately 3 ms at injection, with 100 kV peak gap voltage.

At transition, the main mechanism for beam loss is bucket mismatch and not from space charge [6]. The 2nd harmonic cavity can be used to shape the bucket so that the beam is better matched to it before and after transition. At extraction, the cavity can be used to linearize the voltage during bunch rotation so that there can be a reduction in the tails of the rotated distribution. For more details and references, see [7].

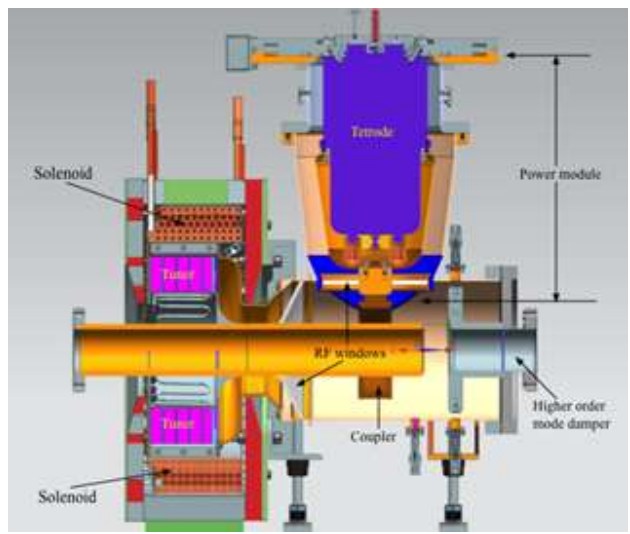

Figure 1: Model of the finalized cavity design. The length is 844 mm from flange to flange.

## CAVITY DESIGN [8,9]

A model of the cavity is shown in Fig. 1. The flange-to-flange length is 844 mm, the aperture is 76 mm, and the shunt impedances are 96 kΩ and 180 kΩ, at 76 MHz and 106 MHz, respectively. The shunt impedance $R_{sh}$ is defined as $V_p^2/2P$, where $P$ is the average dissipated power and $V_p$ is the peak voltage.

The cavity is a quarter wave type and is shorted at the garnet end. The magnetic field in the garnet rings is generated by a solenoid contained within a flux return and two pole pieces. Two alumina windows are used so that the tuner rings and the power amplifier (PA) are outside of the cavity/beamline vacuum. The power amplifier, which uses a cathode driven Eimac Y567B (4CW150000) tetrode, sits between the garnet and the gap end, and is capacitively coupled by a ring which surrounds the cavity's center conductor. The accelerating gap end has a Smythe [10] type higher order mode damper.

Although perpendicularly biased cavities have the advantage that RF losses are lower, the tuning range for this particular cavity is large and there are many technological challenges to overcome. Some of the concerns are 1) Achieving the required tuning range using a realistic bias magnetic field, 2) Keeping the magnetic field in the tuner as uniform as possible (including minimizing the effect of eddy currents), 3) Taking into account higher local permeability and heating of the garnet in areas of lower magnetic field, 4) Including the power amplifier/tetrode in the RF model to take into account the impact of the additional volume and tetrode output capacitance on the cavity tuning range, 5) Choosing a design concept for the tuner that would simplify removal of the heat generated by both the RF and eddy currents, and 6) Finding a safe alternative to using BeO as the heat transfer material.

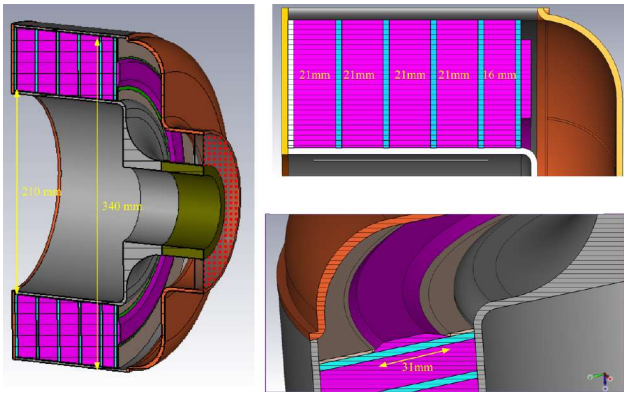

Figure 2: The tuner section of the cavity.

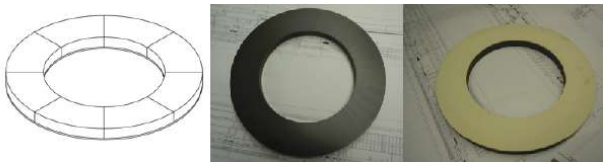

Figure 3: Layout and photographs of the garnet and alumina sides of a tuner ring.

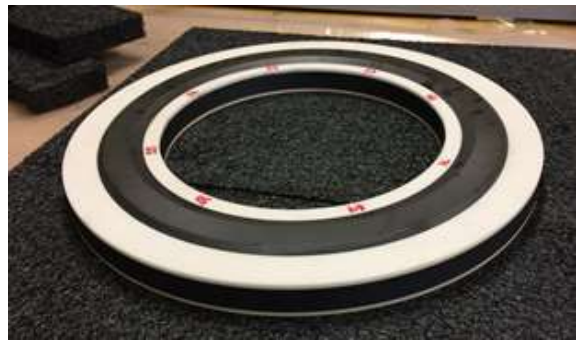

Figure 4: Tuner Shim Ring. The full ring consists of one ring of alumina on the bottom, one garnet ring, a second ring of alumina, and finally the angled shim ring.

## TUNER

The tuner section of the cavity is shown in Fig. 2. Each of the five rings uses National Magnetics AL-800 material, which is aluminium doped garnet. The saturation magnetization ($4\pi M_s$) is 800 G and the Curie temperature is 210 °C. The OD and ID of the rings are 340 mm and 210 mm, respectively, and the thickness is 21 mm. The garnet rings, shown in Fig. 3, are made from eight sectors which are epoxied (Stycast 2850FT, Catalyst 9) together. This allowed the use of an existing oven with a well-known temperature distribution. To remove heat from the garnet material, each ring is epoxied to a 3 mm thick alumina (99.5%) ring. Each garnet sector is cut from a brick along with a small "witness sample", which is to be used for quality control.

Regions of high RF power loss occur where the bias magnetic field approaches the gyromagnetic resonance value, and the power loss in the material increases sharply. The transition area between the garnet and the beampipe is especially vulnerable in this respect due to large changes in the permeability of the media. To improve the field quality in this area, a specially shaped shim ring was added to the front of the tuner stack. A photograph of it is shown in Fig. 4. With the addition of the shim, the minimum value of the magnetic field in the garnet increases from 33 Oe to 69.3 Oe. This is comfortably higher than 27 Oe, the field at which the gyromagnetic resonance occurs for the lowest frequency.

The cavity frequency is tuned by changing the bias magnetic field in the garnet, which changes the permeability in the garnet. The entire Booster ramp occurs during a period of ~ 33 ms. It was necessary to take steps to reduce the eddy currents induced in the walls of the cavity during the ramp, which would generate heat and distort the bias field. The center and outer conductors in the tuner part of the cavity are made from stainless steel, which has poor conductivity. Also, they are divided into four azimuthal sections, which interrupts the azimuthal component of the eddy currents. In order that the stainless steel not cause RF losses, the center and outer conductors are copper plated on the sides which encompass the RF volume.

The RF heat generated in the garnet, which has poor thermal conductivity, is transferred by the alumina disks to the center and outer conductors, which are water cooled. The design is such that the temperature in the garnet does not exceed 100 °C This water cooling also serves to remove heat generated by eddy currents.

### Solenoid and Bias Supply

The bias magnetic field is generated between the two poles of a magnet with solenoidal windings. The fast cycling of the Booster at 15 Hz required that the yoke be made from silicone steel laminations to reduce the effects of eddy currents.

The complete winding is made of two coils wound using 0.46" square copper wire with a 0.25" hole for cooling. The total number of turns is 60. A photograph is shown in Fig. 5. The coils can be powered independently, or connected in series. In the present design the series connection is used. The required current ramps for the cavity to operate at injection, transition, and extraction, are shown in Fig. 6.

Due to funding constraints in FY2018, the complete bias supply has not yet been purchased. The bias supply now consists of four Performance Controls, Inc. GA301-

VP amplifiers in parallel, and will be upgraded in the future with another seven or eight such supplies. So, initially, due to the limited amount of power from the existing supply, the cavity will be operated only for the first 3 ms at the most critical beam loss points, which is during injection/capture.

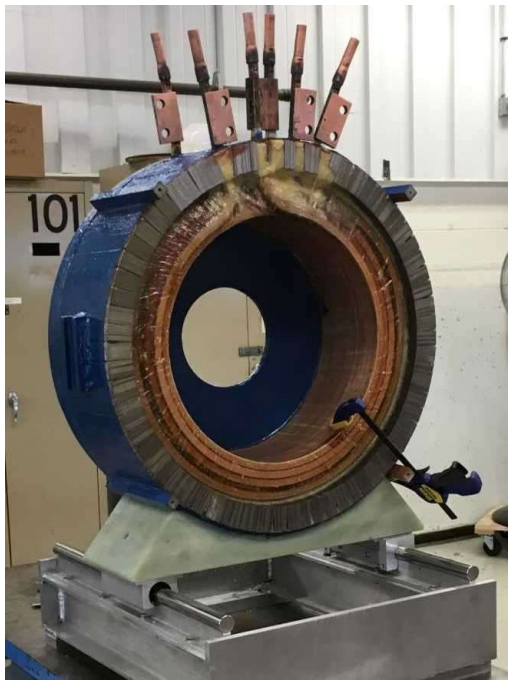

Figure 5: The bias solenoid with the end plates removed.

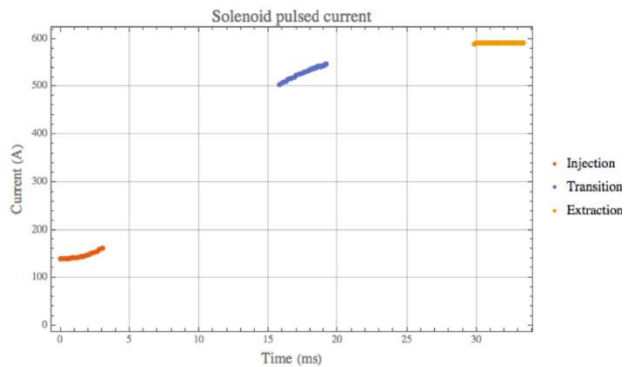

Figure 6: The current pulses required to bias the garnet during the critical periods. The current ramp outside these regions will be determined operationally.

## GARNET CHARACTERIZATION

The accurately model the cavity it was necessary to know the garnet permeability as a function of magnetic field. The real ($\mu'$) and imaginary ($\mu''$) parts of the permeability determine the tuning range and losses, respectively. The magnetic field in the tuner is never perfectly uniform, and to properly model the device, it must be known at every point in the tuner for all bias settings. Since such data is not published, the static permeability and loss tangent were measured as a function of magnetic field using a set of ten AL-800 garnet rings (3.0" OD, 0.65" ID, and 0.5" thick), and a solenoid. A more detailed discussion of these measurements may be found in [11].

### Static Permeability

Ten stacked rings were placed inside the solenoid, which has a ferrite flux return on the bottom and sides. A steel plug was inserted on top to improve the uniformity of the magnetic field within the samples. Three Hall-type magnetic probes were used to measure the magnetic field at the top, bottom, and middle of the stack as a function of solenoid current.

The static magnetization curve was then extracted by iteratively adjusting the magnetization curve used in the simulation of the setup, until the simulation results matched measurements. The initial $\mu(B)$ curve was a guess based on the vendor's data for the initial permeability (~50), and a theoretical value of the permeability well above saturation.

### Loss Tangent

To measure the magnetic loss tangent of the garnet, a quarter wave coaxial test resonator was constructed and filled with the same set of garnet rings that were used for the static permeability measurement. The resonator was placed inside of the solenoid, and the resonant frequency and quality factor $Q$ of the cavity was measured as a function of bias current. As $Q$ is an integrated quantity, and the loss tangent depends on the magnitude of the magnetic field and the frequency, an iterative approach was used (as before) since the field in the sample is not uniform.

Magnetic power losses are traditionally characterized by a loss coefficient $\alpha$. Since $\alpha \ll 1$, the theoretical expression for the loss tangent can be written in a simplified form:

$$\tan \delta_\mu = \frac{\mu''}{\mu'} \approx \frac{\alpha \omega \omega_M (\omega_0^2 + \omega^2)}{(\omega_0^2 - \omega^2)(\omega_0^2 - \omega^2 + \omega_M \omega_0)}$$

where $\omega = 2\pi f$ is the RF frequency, $H_0$ is the magnetic field in the material, $\omega_0 = \mu_0 \gamma H_0$ is the precession frequency, $\gamma = e/m_e$ is the gyromagnetic ratio, $\omega_M = \mu_0 \gamma M_s$, where $M_s$ is the saturation magnetization. For a material with properties parameterized by $\omega_M$ at a point with the field given by $\omega_0$ and at RF frequency $\omega$, the loss tangent is proportional to $\alpha$.

Values of $\alpha$ for the AL-800 were determined for each current setting in the solenoid by adjusting its value in the model until the predicted values of $Q$ and $f$ were the same as seen in the data. It was close to constant, except below ~40 A, where a sharp rise was observed. This can be explained by the onset of gyromagnetic resonance in some (initially small) parts of the sample, an effect which was not seen in the modelling. Therefore, it was concluded that $\alpha$ is constant with a value of 0.0033.

### Tuner Ring Testing

To verify that the magnetic properties of all garnet rings are identical (or close), a test setup was constructed to measure each fully assembled ring. It consists of a test cavity and a bias magnetic system and was designed to ensure that the magnetic field in the garnet was as uniform as possible. The setup of the low power RF cavity and bias system are shown in Fig. 7. Two weakly coupled probes are used to measure the frequency and $Q$. These vary depending on the bias magnetic field in the ring, which is generated by the solenoid. For measurement, the garnet/alumina rings are placed in the large OD section of the cavity at the shorted end.

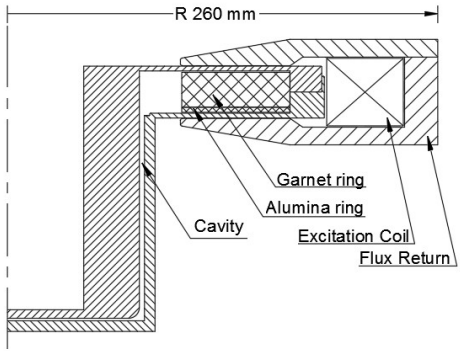

Figure 7: Schematic of the tuner test cavity with magnet.

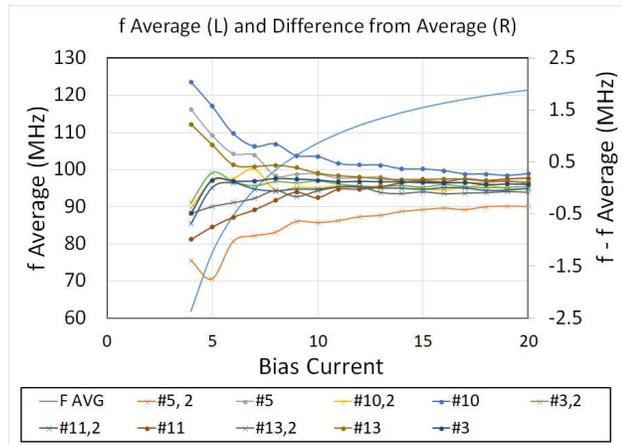

Figure 8: Measured frequency (average and difference from average) for the five tuner rings. The plot shows two measurements for each ring.

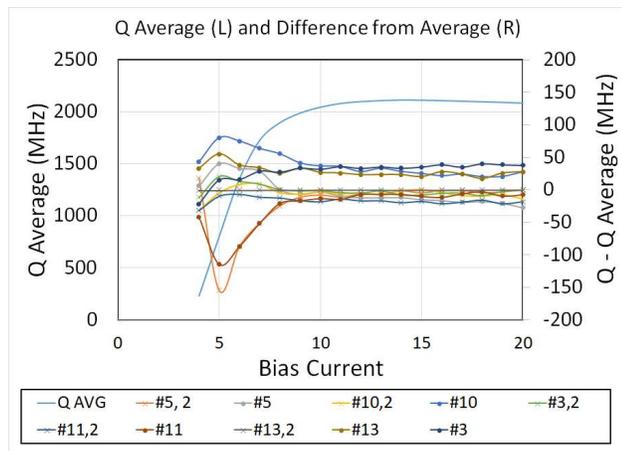

Figure 9: Measured $Q$ (average and difference from average) for the five tuner rings. The plot shows two measurements for each ring.

The biasing coil is made from 224 turns of #10 square copper wire. The flux return is made from 1010 low carbon steel.

Figs. 8 and 9 show the measured resonant frequency and quality factor ($Q$) of the test cavity with each of the five tuner rings. Two measurements are shown for each ring to quantify the repeatability of the measurements. Between successive measurements of the same ring, the cavity top was removed and then reinstalled. The results for the frequency agree well with predictions from simulation (which used sample measurements discussed earlier in this section) and show acceptable scatter, in that the differences between the measured values for one ring are similar to the differences for two different rings. The $Q$ measurements are also uniform from ring to ring, but in this case, it was not possible to match measurement data with simulation. This may be because the loss coefficient α is not constant, although initial measurements on smaller samples seemed to indicate otherwise. Nevertheless, the rings are very uniform in their material properties indicating good quality control by the vendor.

In addition to measuring the properties of the fully assembled rings, the static permeability was measured for each witness sample, corresponding to each sector from each and every ring. The permeability measurements from all witness samples show that they are satisfactorily uniform. This consistency gives confidence that the sectors of the fully assembled garnet rings do not have significant variation in their magnetic properties. More details regarding the witness sample measurements and ring measurements may be found in [12].

## POWER AMPLIFIER TESTS

According to specifications, the Y567B tetrode (Eimac 4CW150000) which will be used in the PA for this cavity can operate up to 108 MHz with 150 kW of power dissipated in the anode. Still, it was desirable to verify this by power testing the tetrode at our operating frequencies. In initial tests at 76 MHz an output power of ~140 kW was obtained (the cavity requires 52 kW at injection). Before power testing at 106 MHz (the extraction frequency) it was necessary to modify the drive part of the PA.

A cathode resonator is used to match the 50 Ω output impedance of the TOMCO 8 kW solid state driver amplifier (SSD) to the tetrode input capacitance. It is a shorted transmission line with inductive impedance opposite that of the tetrode input capacitance, and is resistively loaded to make the resonance very broad to accommodate the ~30 MHz frequency swing. Its dimensions were initially chosen so that the reflected power was minimized at 76 MHz, where the cavity's shunt impedance is lower, but still acceptable at 106 MHz. After constructing the first cathode resonator, it was found that its response at 106 MHz was barely acceptable and not as predicted by the modelling. The trend indicated that the effective input capacitance was frequency dependent. A new model which took this into account was used in the subsequent calculations. The cathode resonator was then redesigned so that the reflected power was minimized at 91 MHz instead of 76 MHz, as in the initial design. Though the cavity will never operate at 91 MHz, this configuration produced acceptable amounts of reflected power at both 76 and 106 MHz.

With the redesigned cathode resonator, the PA was tested at both 76 and 106 MHz. At 76 MHz a quarter wave output (anode) resonator was used. At 106 MHz, a quarter wave resonator would have been impractically small, so a 3/4 wavelength resonator was used. A water cooled 50Ω load is loop-coupled to the anode resonator to absorb the output power, which is measured calorimetrically (by change in water temperature). For several values of DC anode voltage, the drive power was increased until the screen current was 300 mA. We obtained 110 kW and 145 kW maximum output power at 76 MHz and 106 MHz, respectively. Fig. 10 shows various quantities for the 106 MHz test. Details of the PA tests are documented in [13].

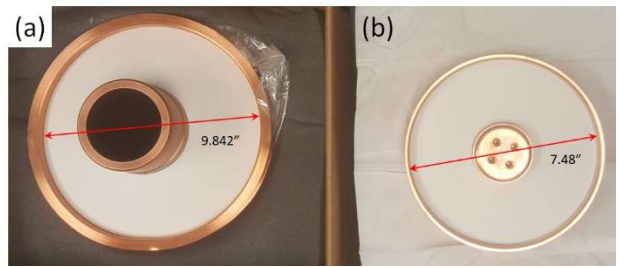

Figure 11: (a) Conical tuner window and (b) flat input window.

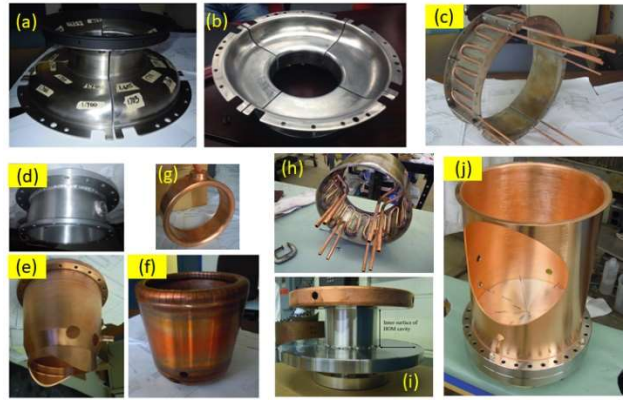

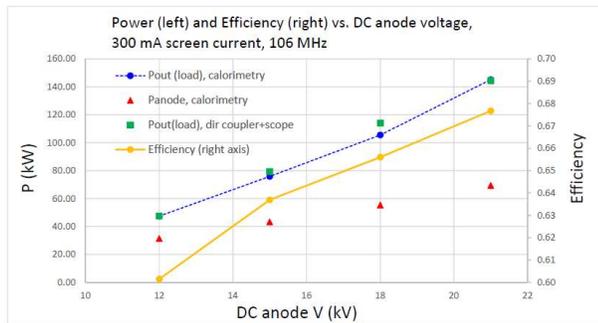

Figure 10: Output power, power dissipated in the anode, and efficiency for Y567B tests at 106 MHz.

Figure 12: Copper and stainless-steel parts of the cavity. (a) and (b): Garnet-free region of the tuner outer conductor. (c): Section of the tuner outer conductor which will contain the garnet. Both parts of the tuner outer conductor are split into four azimuthal sections to reduce eddy currents. (d) and (e): Two outer conductor sections of the PA part of the cavity. They contain ports for airflow, water cooling, monitoring, and the anode DC voltage input. (f): The PA center conductor which supports the tetrode and makes electrical contact with its anode. (g): The ring which capacitively couples the PA to the cavity. (h): The tuner center conductor with cooling lines. (i): The HOM damper cavity. (j): The main cavity outer conductor (vacuum section). The PA section shown in (e) will be attached at the large hole in the OD.

## CONSTRUCTION STATUS

All the cavity parts have been received. The conical and flat windows are shown in Fig. 11. The former is used to isolate the tuner volume, and the latter, the PA volume, from the rest of the cavity. This is so any outgassing materials used in the tuner construction do not contaminate the vacuum space, and so a PA may be changed without breaking the beam/cavity vacuum.

Other critical parts are shown in Fig. 12. At the time of this writing, the PA window has just been welded into the cavity, and the PA outer conductor has been welded on. This is shown in Fig. 13.

## CONCLUSION

We have designed and are in the process of constructing a 2nd harmonic cavity for the Fermilab Booster. We have measured the magnetic properties of the fully assembled tuner rings and their corresponding witness samples. The permeability and losses in the material are very uniform.

The cavity will be tested early this summer and installed into the Booster during a planned shutdown. If successful, this will be the first operational broadband perpendicularly biased cavity and it will be a significant technical achievement for accelerators.

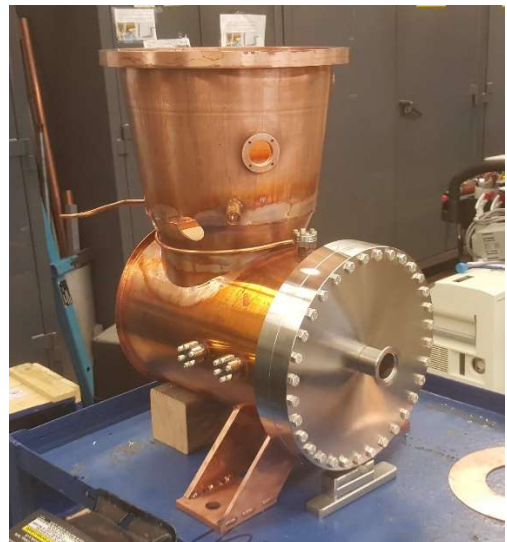

Figure 13 : The partially assembled cavity.

## ACKNOWLEDGMENTS

Many thanks to National Magnetics for the manufacture of our garnet and the assembly of the tuner rings with our required consistency and precision.